\documentclass[%
 reprint,
 amsmath,amssymb,
 aps, 
prb,
]{revtex4-2}

\DeclareUnicodeCharacter{2215}{/}

\usepackage{morefloats}
\usepackage{color}
\usepackage{epsfig,graphicx,amsfonts,amsbsy}
\usepackage{amsmath,amsfonts,amsthm,amssymb}
\usepackage{appendix}
\usepackage{bbm}
\usepackage{makeidx}
\usepackage{url}
\usepackage{verbatim}
\usepackage{mathrsfs} 
\usepackage{morefloats}
\usepackage{appendix}
\usepackage{bbm}
\usepackage{makeidx}
\usepackage{url}
\usepackage{verbatim}
\usepackage[bookmarksnumbered,pdfpagelabels=true,plainpages=false,colorlinks=true,linkcolor=blue,citecolor=blue,urlcolor=blue]{hyperref}
\usepackage[bookmarksnumbered,pdfpagelabels=true,plainpages=false,colorlinks=true,linkcolor=blue,citecolor=blue,urlcolor=blue]{hyperref}
\usepackage{array}
\usepackage{booktabs}
\usepackage{multirow}
\usepackage{bbm}
\usepackage{tabularx}
\usepackage{cancel,soul}
\usepackage{nicefrac, xfrac}
\usepackage{ulem}
\usepackage{bbold}

\newcommand{\bS}{\boldsymbol{{\rm S}}}

\newcommand{\bpsi}{\boldsymbol{{\rm \psi}}}

\newcommand{\bmm}{\boldsymbol{{\rm m}}}
\newcommand{\bn}{\boldsymbol{{\rm n}}}
\newcommand{\bN}{\boldsymbol{{\rm N}}}
\newcommand{\bB}{\boldsymbol{{\rm H}}}

\newcommand{\cH}{{{\cal H}}}
\newcommand{\Jint}{J_{\text{int}}}

\newcommand{\rvec}{\mathbf{r}}

\newcommand{\sectioncustom}[1]{\section{#1}}
\newcommand{\subsectioncustom}[1]{\subsection{#1}}

\def \fcfm {Departamento de F{\'i}sica, FCFM, Universidad de Chile, Santiago 8370449, Chile.}

\begin{document}


\title{Electrical Control of the Exchange Bias Effect at Model Ferromagnet-Altermagnet Junctions}

\author{Gaspar De la Barrera}
\email{gaspar.delabarrera@ug.uchile.cl}
\author{Alvaro S. Nunez}
\email{alnunez@dfi.uchile.cl}
\affiliation{\fcfm}
\date{\today}

\begin{abstract}
This work analyzes the behavior of the interface between a ferromagnetic material and an altermagnet. We use a well-established line of arguments based on electronic mean-field calculations to show that new surface phenomena that lead to altermagnetic materials induce an exchange bias effect on the nearby ferromagnet. We reveal the physical mechanisms behind this phenomenon that lead to quantitative control over its strength.  Interestingly, we predict exotic electric field-induced phenomena. This is an analogy to the relationship between exchange bias and the injection of spin currents in spin-transfer-dominated scenarios, which has been reported earlier in the traditional antiferromagnetic/ferromagnetic junction.
\end{abstract}

\maketitle








\sectioncustom{Introduction}
Recent advances have introduced a novel type of collinear magnetism referred to as altermagnetism\cite{zhou2024, reimers2024, krempasky2024, fedchenko2024, Bai2024, Hajlaoui2024}. Such a magnetic phase features a strong breaking of time-reversal symmetry, a staggering magnetic configuration, and spin-splitting in its band structures while maintaining zero net magnetization. Altermagnetism integrates attributes once thought exclusive to conventional collinear ferromagnetism or antiferromagnetism, enabling new phenomena and capabilities unheard of within these traditional magnetic categories. Initially suggested through theoretical models, the existence of the altermagnetic phase has been confirmed by various experimental studies that have validated its distinctive characteristics and potential for practical applications.

 In altermagnetic materials, the magnetic moments (spins) align, leading to a pattern of alternating magnetization on an atomic scale. However, unlike antiferromagnets, the electronic band structure remains split by spin\cite{Hayami2019, Hayami2020}. \textcolor{black}{
This is a direct consequence of spin-group symmetry [9,10], which leads to an electronic structure breaking Kramers degeneracy, even in the absence of an external magnetic field. This
leads } to unusual transport properties\cite{Smejkal2022Landscape}, piezomagnetism \cite{Aoyama2024}, the generation of spin-splitter torque in MRAM geometries\cite{Karube2022}, chiral split magnon bands\cite{Smejkal2023, Liu2024}, and as a possible route towards Majorana physics\cite{Ghorashi2024}, among others. Recently, it was shown that nanoscale imaging of altermagnetic states in manganese telluride (MnTe) is possible\cite{Amin2024}. 
Several theoretical works have focused on modulation and realization of \textcolor{black}{altermagnetic phases in materials} \cite{Gomonay2024, Smejkal2023, Ferrari2024, Durrangel2024}, as well as the generalization of altermagnetism to noncollinear magnets\cite{Cheong2024}. Furthermore, the anomalous Hall effect was reported in altermagnetic rutenium dioxide\cite{Feng2022}.

This concept is new in condensed matter physics and is currently a subject of active research as scientists explore its fundamental properties and potential technological applications
\cite{Leeb2024}. In addition, we highlight a profound connection between altermagnetism and the physics of superfluid He$^3$ \cite{Jungwirth2024} that promises a myriad of analogies.

The role altermagnetism will play in heterostructures is still in the making. Some works point to exciting effects in superconducting-altermagnetic hybrids, where effects such as nonreciprocal superconductivity, also known as the superconducting diode effect, have been predicted\cite{Banerjee2024}. 

In the field of altermagnetic spintronics, there are proposals for giant magnetoresistance and tunnel magnetoresistance\cite{Smejkal2022, Chen2024b}. These are critical physical effects used in commercial spintronics devices to read information. These effects are based on the transmission of a spin current between a reference ferromagnetic electrode and a sensing one within a multilayer structure. Naturally, they can occur in the context of altermagnetic systems. Spin Hall and Edelstein effects in noncollinear altermagnets were predicted in\cite{Hu2024}. Recently, spin transfer phenomena have been discussed in the context of altermagnetism \cite{Zarzuela2024}. In this work, we focus on the effect of exchange bias.

 Exchange bias (EB) is a phenomenon observed in systems consisting of an interface between a ferromagnetic material (FM) and an antiferromagnetic material (AF) \cite{isingTypeAF, Blachowicz2023-bd, nogues1999exchange, nogues2005,lin2019,kohn2013}, which results in a shift of the material's magnetization curve when exposed to an external magnetic field. This phenomenon, first discovered in the 1950s, has been the subject of extensive research due to its relevance in technological applications, such as magnetic storage devices\cite{Sharma}. Today, it is part of the standard lore of magnetism research and is relevant in modern low-dimensional magnetism research\cite{zhu2020, Hasan2023, chen2024, bao2024, cham2024}. The exchange bias effect has been observed in multiferroic field effect devices\cite{Wu2010} as well as polycrystalline thin films\cite{OGrady2010}. The picture behind the physical processes involved in the effect of exchange bias has become more apparent since the different roles played by the nucleation and propagation of domain walls in exchange bias were disentangled in a recent article\cite{Hasan2024}. 
 The theory of EB has evolved over the years and several models have been proposed to explain the effect\cite{kiwi2001exchange, Stamps2000}.

The above leads us to question how the interface between a ferromagnet and an altermagnet would behave. This could generate interesting emergent interactions because of altermagnetism (AM) similarity to antiferromagnetism (AF). Still, the additional properties of altermagnetism could result in different behaviors, such as spin density oscillations in AM and control of AM dynamics by an electric field\cite{Wang2024}. Moreover, it opens up the possibility of working with conducting materials since the relationship between Exchange Bias (EB) and the injection of electric currents has been studied \cite{tesisAlvaro, MacDonald2011, Kim2019}. This work analyzes the behavior of an interface between a ferromagnetic material and an altermagnetic material using all-fermion calculations. We use a line of arguments along \cite{Weber2024} and show that new phenomena occur because of a material with spin-split bands.

\sectioncustom{Basic model}\label{sec: model}
The basic altermagnetic/ferromagnet (AM/FM) hybrid model is based on the model described in \cite{Leeb2024, Solovyev2024}. We consider a square lattice. A two-orbital electronic system with Hamiltonian written in terms of fermion operators $\bpsi_{\mathbf{r} a\sigma}$, where the label $\rvec$ stands for space position, the label $a$ for orbital and the label $\sigma$ for spin species.
The kinetic energy is written as\cite{Raghu2008}: 
\begin{equation}
    \cH_t=-\sum_{\langle \rvec, \rvec'\rangle}\sum_{ab} \sum_{\sigma}t^{ab}_{\rvec-\rvec'}\bpsi^\dagger_{\rvec a\sigma}
\bpsi^{\phantom\dagger}_{\rvec'b\sigma}.
\end{equation}

\begin{figure}
    \centering
    \includegraphics[width=1\linewidth]{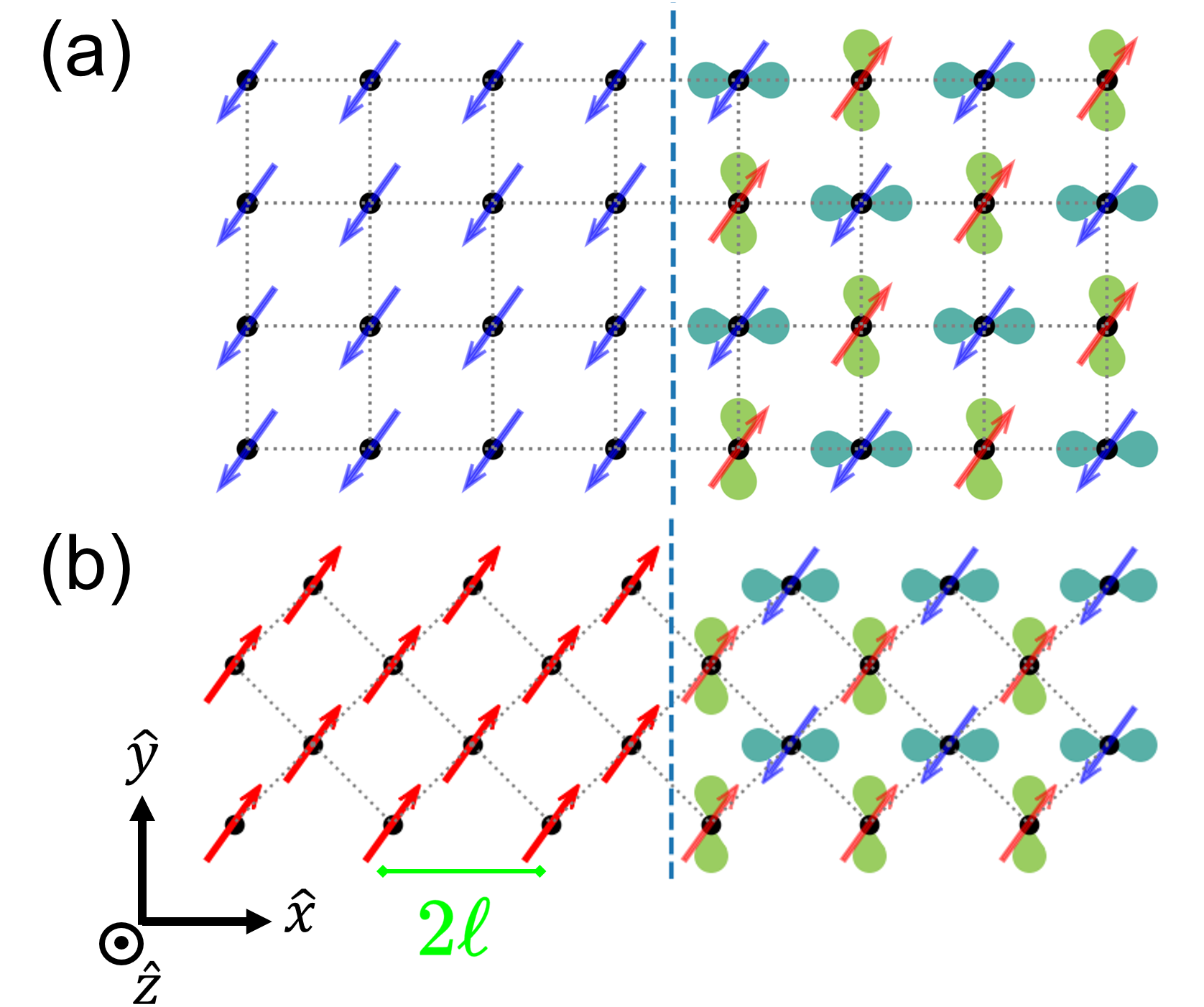}
    \caption{Cartoon of the model described in the main text for (a) a square lattice, with compensated surface; and (b) a centered square lattice, that's just a rotation of a square lattice, with uncompensated surface. The colored dot represents the spin component, blue meaning up and red down. The orbitals are depicted by orthogonal lobes that alternate (in the AM region). At the interface, the last row of the altermagnet is commensurate with the first row of the altermagnet. Their spin coupling is defined to be $\Jint$. The deficit in the coordination of the last altermagnetic row results in a magnetically weak layer. The distance \textcolor{black}{$\ell$ will be used as a unit of length through this paper and it characterizes the scale of the system.} }
    \label{fig: cartoon}
\end{figure}

The coefficients $t^{ab}_{\rvec - \rvec'}$ are obtained from a discretization of the Hamiltonian in \cite{Leeb2024}. \textcolor{black}{Explicitly, this matrix contains hoppings to nearest and next-nearest neighbours, depending on the orbitals. In momentum space it is,}

\begin{equation}
\hat{t}_\mathbf{k} = \begin{pmatrix}
    \varepsilon_x (\mathbf{k}) & \varepsilon_{xy} (\mathbf{k})\\
    \varepsilon_{xy} (\mathbf{k}) & \varepsilon_y (\mathbf{k})
\end{pmatrix}
\end{equation}

where
\begin{align}
        \varepsilon_x(\mathbf{k}) &= -2t_1 \cos k_x - 2 t_2 \cos k_y - 4 t_3 \cos k_x \cos k_y \\
        \varepsilon_y (\mathbf{k}) &= -2t_2 \cos k_x - 2 t_1 \cos k_y - 4 t_3 \cos k_x \cos k_y \\
        \varepsilon_{xy}(\mathbf{k}) &= -4 t_4 \sin k_x \sin k_y \,,
\end{align}
\textcolor{black}{are the energies associated with the electronic transport between the two possible orbitals, $x$ and $y$, at each site. The values of the hopping parameters depend on the material, the ones used in this study are the same as in \cite{Leeb2024}: $t_1 = -t$, $t_2 = -1.75t$, $t_3 = -0.85t$ and $t_4 = -0.65 t$. It should be noted that these values are for a square lattice, showed in the right hand side of Fig. (\ref{fig: cartoon}a), we have to do a $45^\circ$ rotation to the $k_x$ and $k_y$ values.}

Additionally, the model is supplemented by two interaction energies, a Heisenberg-like exchange in spin space, $\cH_J$, and an Ising-like exchange in orbital space, $\cH_V$. These contributions are given by:
\begin{equation}
\cH_J=J \sum_{\langle \rvec, \rvec'\rangle} \bS_\rvec \cdot\bS_{\rvec'},
\mbox{    and     }
\cH_V=V \sum_{\langle \rvec,\rvec'\rangle} \bN^z_{\rvec}\,\bN^z_{\rvec'},
\end{equation}
where $\bS_\rvec=\sum_{a, \sigma,\sigma^\prime}\bpsi^\dagger_{\rvec a\sigma}\vec{\tau}_{\sigma\sigma^\prime}\bpsi^{\phantom\dagger}_{\rvec a\sigma^\prime}$ and $\bN_i=\sum_{a,b,\sigma}\bpsi^\dagger_{ia\sigma}\vec{\tau}_{ab}\bpsi^{\phantom\dagger}_{ib\sigma}$, \textcolor{black}{where $\vec{\tau}_{ab}$ are the usual components of the Pauli matrices}. For positive $J$ and $V$, we therefore expect that $H_J$ induces $(\pi, \pi)$-AFM (antiferromagnetic order) and, crucially, that $H_V$ induces $(\pi, \pi)$-OO (orbital order, \textcolor{black}{sometimes called antiferro-orbital order}).  Due to the absence of spin-orbit coupling, $[\cH, \bS_\rvec] = 0$, this implies that the spin remains a good quantum number in the original model. 
\textcolor{black}{
We will study the effect of applying and electric field to the system, so we include the following term:
}

\begin{equation}
\cH_\Delta=  \sum_{\rvec a\sigma} \Delta_\rvec  \bpsi^\dagger_{\rvec a\sigma}\bpsi^{\phantom\dagger}_{\rvec a\sigma},
\end{equation}
where $\Delta_\rvec $ stands for a local potential, we have $\Delta_\rvec =-e \vec{E}\cdot(r_x, r_y, (-1)^\rvec )\ell$ where $\vec{E}=(E_x,E_y,E_z)$ and the later term represents an electric field in the out-of-plane direction\cite{Wang2024}. \textcolor{black}{The origin of this term is due to the zig-zag out-of-plane configuration of the crystal. The $\ell$ represents the unit of length of the crystal. The $x$ and $y$ terms of this potential corresponds to the usual of a constant electric field in the corresponding direction.}


Finally, we will drive the magnetic reversal using an external magnetic field $\bB$,
\begin{equation}
\cH_B=-g\mu_0\sum_i \bS_\rvec \cdot\bB.
\end{equation}

The complete Hamiltonian acquires the form: $\cH=\cH_t+\cH_J+\cH_V+\cH_\Delta+\cH_B$. This constitutes the essence of the model. Since we consider an AM/FM junction, the values of the coupling constants will change according to the region. \textcolor{black}{In the ferromagnetic region $J = -J_{\rm FM}$ and in the altermagnetic region $J=J_{\rm AM}$, with $J_{\rm FM}, J_{\rm AM} > 0$. $V_{\rm F}$ will be set to zero and $V_{\rm AM}>0$ will lead to $(\pi,\pi)$-OO.} The interface between the two systems will be described by a different exchange coupling $\Jint$. We make the assumption $J_{\rm FM}< J_{\rm AM}$ while varying $\Jint$ across the parameter space. This parameter will prove essential for the physical behavior of the interface. A cartoon of the model is presented in Fig. (\ref{fig: cartoon}). The deficit in the number of neighbors of the last altermagnetic row results in a magnetically tunable layer whose magnetization will fluctuate as it is subjected to external fields.
We use self-consistent mean-field theory for both the magnetic and orbital order parameters, following the ideas of \cite{Leeb2024}, to determine the local magnetic and orbital orderings for various external magnetic field values. We perform this calculation on a path that simulates a hysteresis cycle.

\sectioncustom{Results} Our results are organized into three paragraphs. The first is a detailed scrutiny of the stability of the altermagnetic order exposed to a magnetic field. The second is an analysis of the interfacial exchange induced by the altermagnet on a nearby ferromagnet. Finally, the third addresses the AM/FM junction exchange bias effect. 

\begin{figure}[h!]
    \centering \includegraphics[width=1\linewidth]{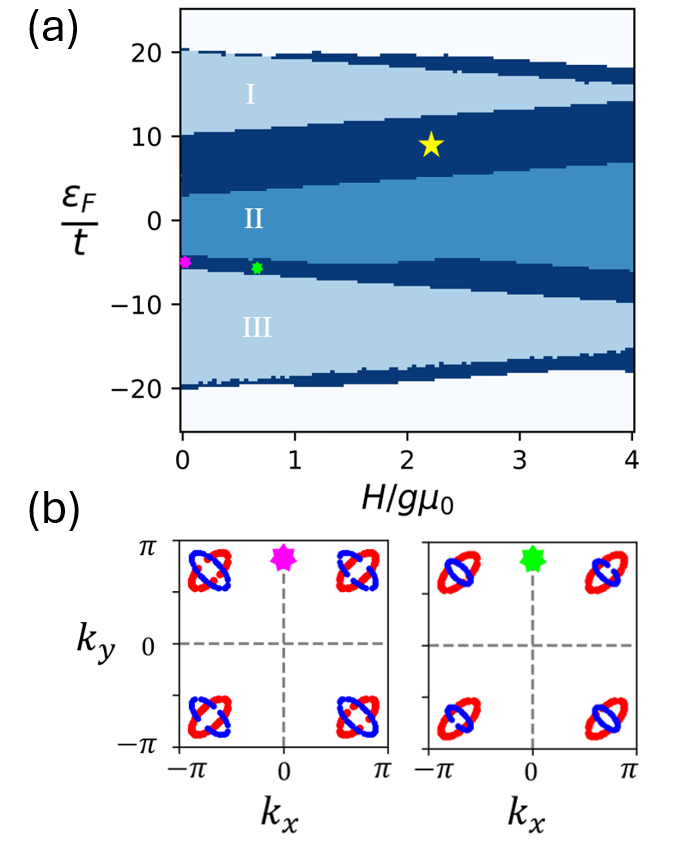}
    \caption{
    Phase diagram of the altermagnet (a) denoting three major regions, (I) OO + AF with $3/4$-filling, (II) OO with half-filling and (III) with OO + AF and $1/4$-filling. The altermagnetic phases (I) and (III) are surrounded by a conducting state marked by the dark blue sections, but for the Fermi energies of the light blue section the altermagnet remains an insulator. The band structure at the fermi level is showed in (b), showing the altermagnetic behaviour for the points denoted in (a), i.e., with and without magnetic field applied. \textcolor{black}{These bands are for the centered square lattice Fig. (\ref{fig: cartoon}b), to obtain the bands for the square lattice Fig. (\ref{fig: cartoon}b) just apply a $-45$° rotation to the $k_x, k_y$ plane. All this is for $J=V=3t$.}
    }
    \label{fig: phase diagram}
\end{figure}

\subsectioncustom{Altermagnetic state vs magnetic field} In this paragraph, we report on altermagnetic stability under the influence of a magnetic field. This analysis will help us to understand the results in subsequent sections. Our calculations are a relaxation of the mean-field equations for different Fermi levels and external magnetic fields. These relaxations are performed for an isolated altermagnet under periodic boundary conditions. \textcolor{black}{We define altermagnetic order as the coexistence of both antiferromagnetic ordering (AF) and antiferro-orbital ordering (OO).} Our results, rendered in Fig.(\ref{fig: phase diagram}), indicate that for a wide range of magnetic fields, the altermagnetic order is stable. Only for large fields is the altermagnetic state challenged. This conclusion is meaningful since it allows us to swipe the field within the hysteresis loop with confidence that the altermagnetic order will be preserved. Regarding the effect of the Fermi level, we must mention that the altermagnetic state is stable in regions where the system is metallic.

\begin{figure}
    \centering    \includegraphics[width=1\linewidth]{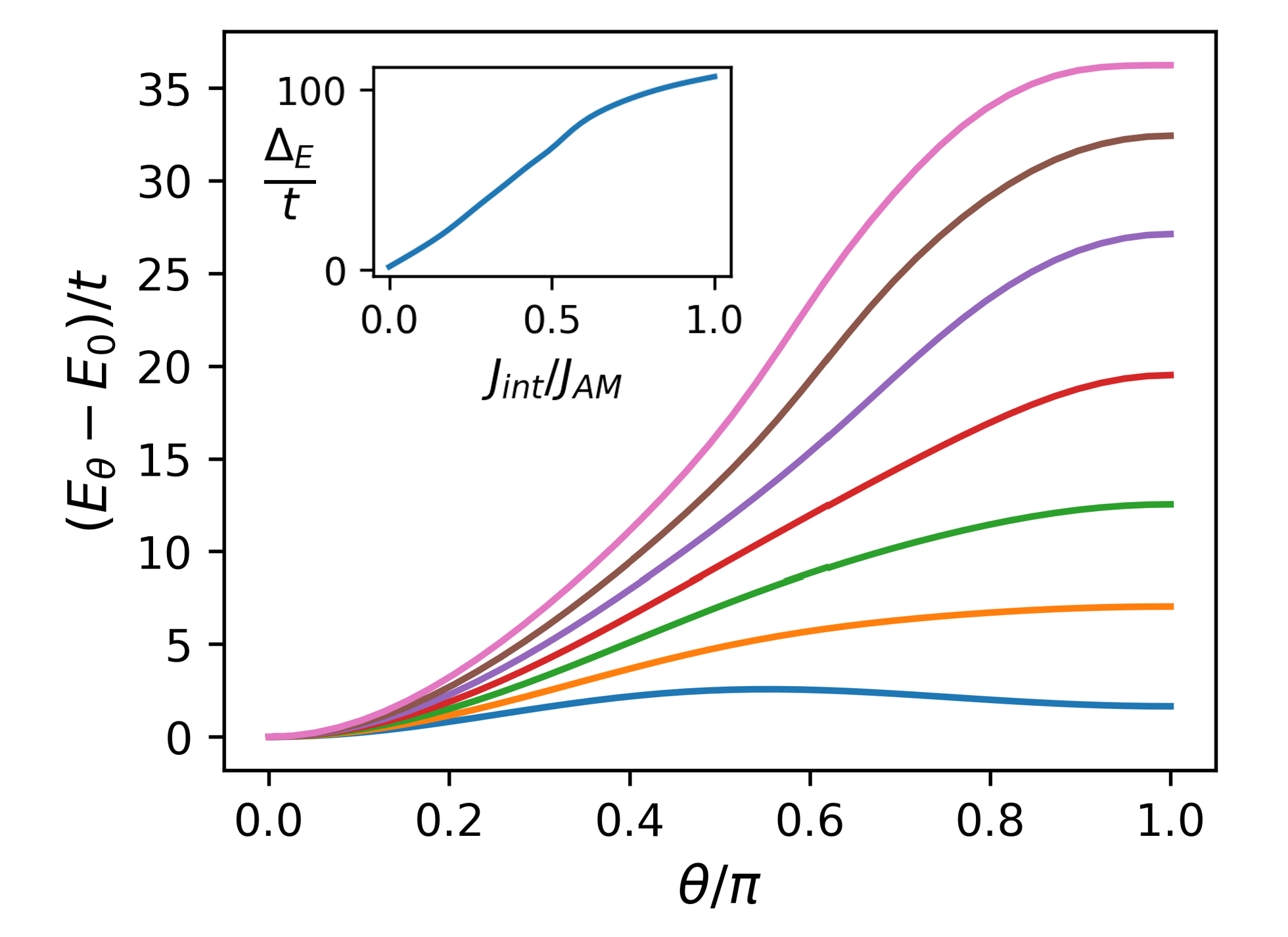}
    \caption{Interfacial exchange interaction with a magnetically uncompensated altermagnetic interface. Main plot: Energy as a function of the relative orientation between the magnetization and the altermagnetic order parameter for seven evenly distributed $\Jint/J_{AM}$ between 0 and 0.3. 
    As shown in the inset, the interfacial exchange strength depends on the interfacial exchange constant $\Jint$. The inset shows the energy gap between orientations, described by $\Delta_E =  E_{\theta=\pi}-E_{\theta=0}$ and its dependency on the interface exchange.}
    \label{fig: anisotropia}
\end{figure}

\subsectioncustom{Interfacial exchange interaction } Here, we study the interfacial exchange interaction between the altermagnet in a ferromagnetic neighbor.
We calculate the net energy of the system prepared with different orientations for the self-consistent spin field.
The strategy is similar to the one in \cite{Nunez2006, Haney2007a, Haney2007b}. We pick an orientation \textcolor{black}{for the antiferromagnetic order parameter of the AM $\mathbf{n} = \sum_{\rvec\in \rm AM} (-1)^\rvec \langle \mathbf{S}_\rvec\rangle/N$} and report the energy associated with the configuration.  
The results are presented in Fig. (\ref{fig: anisotropia}a). Fig.(\ref{fig: anisotropia}b) shows the magnitude of the interfacial exchange interaction as a function of $\Jint$. We appreciate that while for large $\Jint$, the behavior is close to a $\cos\theta$ function (reflecting the usual exchange $\bn\cdot\bmm$, where $\bmm$ is the magnetization vector of the FM), as $\Jint$ is reduced, there is an apparent deviation. In the extreme case $\Jint=0$, the magnitude of interfacial exchange is depressed. This effect is dominated by second-order hopping exchange mechanisms of \cite{Auerbach1994}.  

\subsectioncustom{Exchange Bias}
In this paragraph, we study the effect of exchange bias on the AM/FM interface. This effect manifests itself as a shift in the magnetic hysteresis loop along the axis of the applied field. Specifically, the external field required to reverse the magnetization of the ferromagnetic layer becomes asymmetric with respect to the origin, resulting in a displacement of the hysteresis loop. The exploration of EB and quantum effects in magnetic interfaces opens up a vast and promising field of research. As our understanding of these phenomena deepens and evolves, new opportunities will likely emerge for developing advanced technologies. Therefore, the primary purpose of this section is to investigate the impact of altermagnetism on the EB phenomenon. When the AM/FM system is cooled below the altermagnetic critical temperature while under the influence of an applied magnetic field, the antiferromagnetic moments align and exert an interfacial coupling force on the ferromagnetic layer. This interaction shifts the hysteresis loop in one direction, generating the exchange bias effect.
The term "compensated" is used to refer to an antiferromagnetic (AFM) layer where the magnetic moments at the interface with a ferromagnetic (FM) layer essentially cancel each other out, as in Fig. (\ref{fig: cartoon}a), resulting in a negligible net magnetic moment. At the same time, "uncompensated" means that there are unpaired spins at the interface, as in Fig. (\ref{fig: cartoon}b), leading to a significant net magnetic moment, which is the primary driver of the exchange bias effect in most systems\cite{Cheon2007}.
The same categories can be applied to AM interfaces. 
\textcolor{black}{
In this work, we tried two limit cases, a fully uncompensated structure, where we study the exchange bias and the effect of applying an electric field on the system. Then we study a fully compensated surface (in this case, more realistic systems might achieve different exchange bias by appealing to the extrinsic features not included in our model, such as Dzyalonshinskii-Moriya interactions\cite{Dong2009}). Although idealistic, these models gives an idea of how EB will behave in well constructed epitaxial samples or samples with local domains of compensated/uncompensated surfaces. 
}

\begin{figure}[h!]
    \centering
    \includegraphics[width=
    \linewidth]{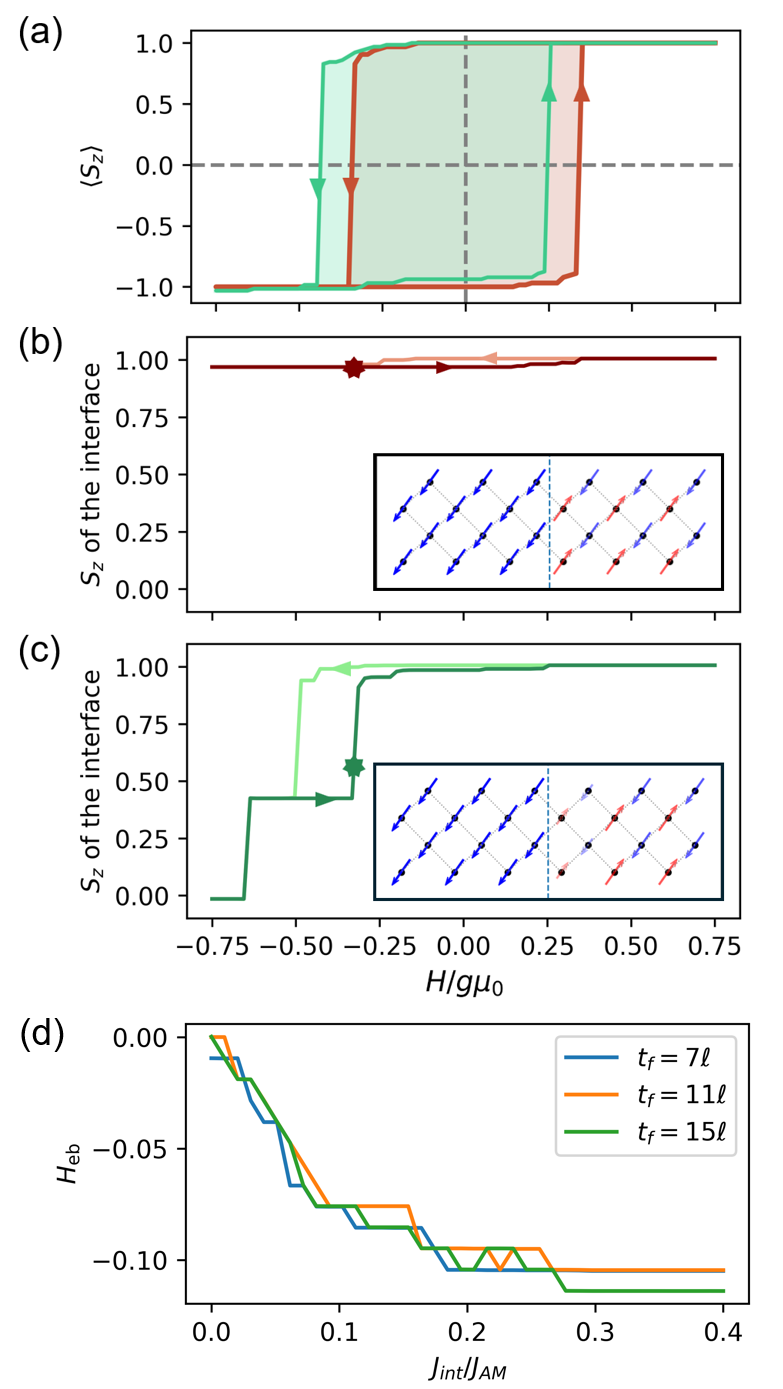}
    \caption{(a) Representative hysteresis cycle showing exchange bias for different interface exchange $\Jint$. The red curve corresponds to $\Jint=0$ and does not display exchange bias as expected. The reversion curves are symmetric for positive and negative fields. The green curve, on the other hand, represents the hysteresis loop for $\Jint=0.2 J_{\rm AM}$. It presents a noticeable exchange bias effect. (b) and (c) Track of the z component of the spin of the AM at the interface along the hysteresis loop. In (b) $\Jint=0$ while in (c) $\Jint=0.2 J_{\rm AM}$ highlights the asymmetric magnetization at the moments of the ferromagnet reversion. In (b) and (c), the inset represents the distribution of spins near the interface in the field indicated by the star. (d) Strength of the exchange bias effect as a function of $\Jint$ for different values of $t_{\rm FM}$, \textcolor{black}{where $t_{\rm FM}$ gives a measure of how big is a system and how many sites there are.}}
    \label{fig: Exchange Bias}
\end{figure}

 We performed fermionic numerical simulations within the mean-field approximation of hysteresis loops for different parameters. The loops are characterized by two reversions at two different coercitivities $H^1_{\rm c}$ and $H^2_{\rm c}$. The exchange bias is characterized by the midpoint between these two: $H_{\rm eb}=(H^1_{\rm c}+H^2_{\rm c})/2$. The reversal process is catalyzed by the nucleation of domain walls at the AM/FM interface. These propagate in response to the external magnetic field crossing the FM, leading to its magnetization reversal. Analysis of the coercitivity fields at the reversal stage allows us to estimate the exchange bias. Through Fig. (\ref{fig: Exchange Bias}), we report a significant strength of the exchange bias effect. In particular, it depends on the interfacial exchange coupling, $\Jint$. For $\Jint=0$, the effects vanish, and as we increase its value, so does the shift in the hysteresis loop.

The spin orientation at the interfacial layer is volatile since it is connected to only half of the neighbors within the altermagnet. In Figs.(\ref{fig: Exchange Bias}b) and (\ref{fig: Exchange Bias}c), we depict the net spin of the interfacial layer along the hysteresis loop. In Fig. (\ref{fig: Exchange Bias}b), $\Jint=0$, and the system presents a vanishing exchange bias effect, while in (\ref{fig: Exchange Bias}c), the $\Jint=0.2 J_{\rm AM}$ is chosen to ensure an exchange bias. The asymmetry of Fig. (\ref{fig: Exchange Bias}c) can be linked directly to the exchange bias effect in the system at hand. 


\subsectioncustom{Effects of an electric field}
\textcolor{black}{
We performed simulations of the whole hysteresis cycle to obtain some important phenomenological parameters of the reversal process (like valid $J$  and $V$ values) and to identify that the reversal process is conducted through a domain wall. To study the dependence of the exchange bias on external electric fields we use the domain wall nucleation model \cite{ZhangDW}, proposed in \cite{cowburn1997multijump}. We numerically calculate the energy of the two metastable states, the one with the ferromagnet polarized in the up direction ($E_\uparrow)$ and in the down direction ($E_\downarrow)$ and determine the reversing field $H_{\rm c}^1$ or $H_{\rm c}^2$ when the difference in energies between both states $\Delta E$ (equal to $E_\uparrow-E_\downarrow$ or $E_\downarrow-E_\uparrow$, depending if one wants to calculate $H_{\rm c}^1$ or $H_{\rm c}^2$) is bigger than a phenomenological parameter $\varepsilon_{180^\circ}$. This parameter represents the energy necessary to nucleate or propagate a $180^\circ$ domain wall, ie, a domain wall that completely turns around the magnetization of the ferromagnet, this parameter depends typically on the sample used, being affected by impurities and other artifacts.
}

\begin{figure}
    \centering
    \includegraphics[width=0.9\linewidth]{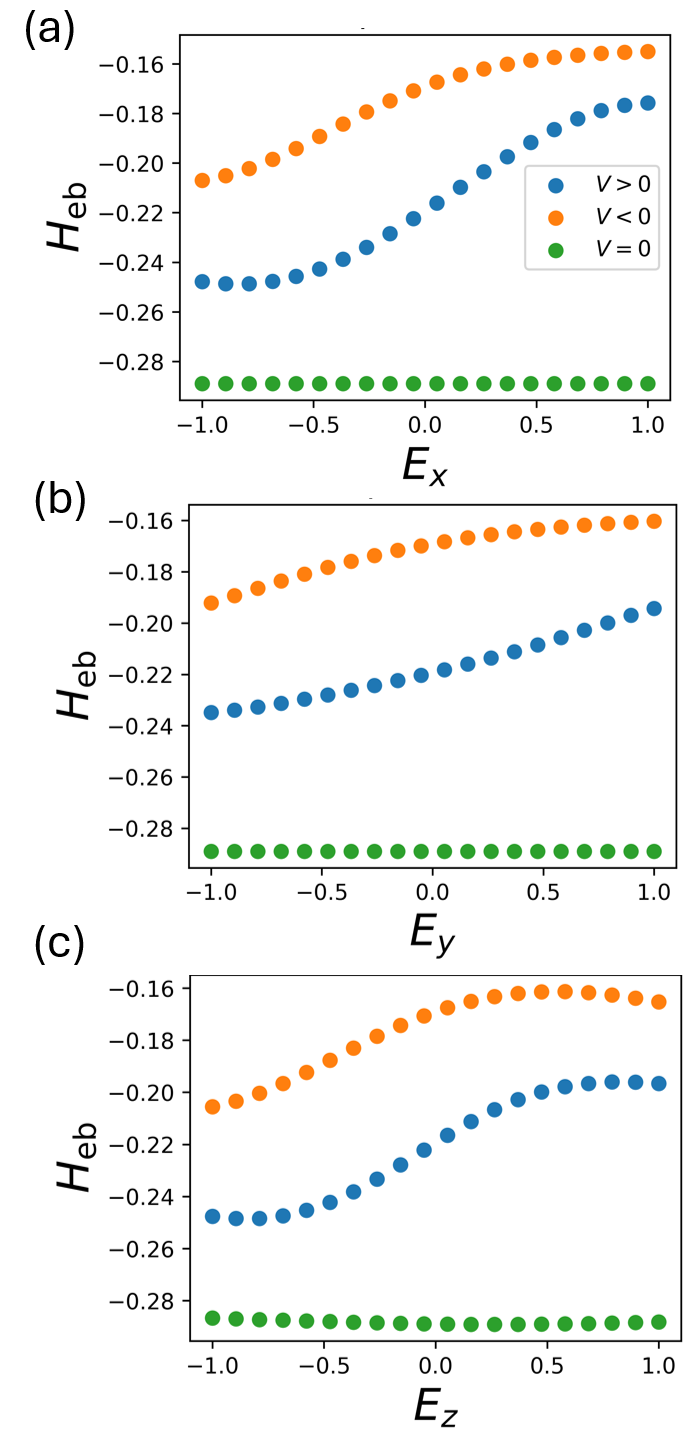}
    \caption{Exchange bias dependence on applying the electric field in the three possible directions. The fields are depicted in $t/(e\ell)$ units. Simulations show a different behavior for the AM ($V>0$), the ferro-orbital-ordered ($V<0$)  and para-orbital-ordered ($V=0$) materials. For the $V\neq0$ cases, $|V| = J = 3t$.}
    \label{fig:campo-electrico}
\end{figure}

The results in Fig.(\ref{fig:campo-electrico}) show clearly the dependence of the exchange bias field on the strength and direction of the electric field.  
\textcolor{black}{
We compare an AM that presents antiferro-ordering in the spin and orbital spaces with two other options that better resemble a usual antiferromagnet. These are the cases with $V=0$, which corresponds to a para-orbital ordering, and $V<0$, which corresponds to ferro-orbital ordering. The bigger $H_{\rm eb}$ for the $V=0$ case showed in Fig. (\ref{fig:campo-electrico}) is due to the para-orbital case, which favors the configuration with both orbitals occupied by the same spin orientation, thus half-filled. This causes the interface to double the magnetization of the 3/4-filled cases: $V>0$ and $V<0$. In all these cases, $V$ denotes the strength parameter of $\cH_V$ in the non-ferromagnetic region.
}

\subsectioncustom{\textcolor{black}{Nature of the Exchange Bias}}

\textcolor{black}{In Fig. (\ref{fig:campo-electrico}), we saw that both the ferro and antiferro-orbital ordering have qualitatively similar responses to the electric field, but they vary in magnitude. So there are differences between the exchange bias of an AM and with other AFM like materials. To better examine this difference and argue that is due to the spin-split nature of the altermagnet, we calculate the exchange bias of both systems with a compensated surface. In this simple model, there are no other anisotropy-inducing mechanisms like in usual realistic exchange bias systems (there are no impurities, magnetic domains, or cooling fields), so one would expect that the exchange bias of a compensated surface would be zero. 
}

\begin{figure}
    \centering
    \includegraphics[width=1\linewidth]{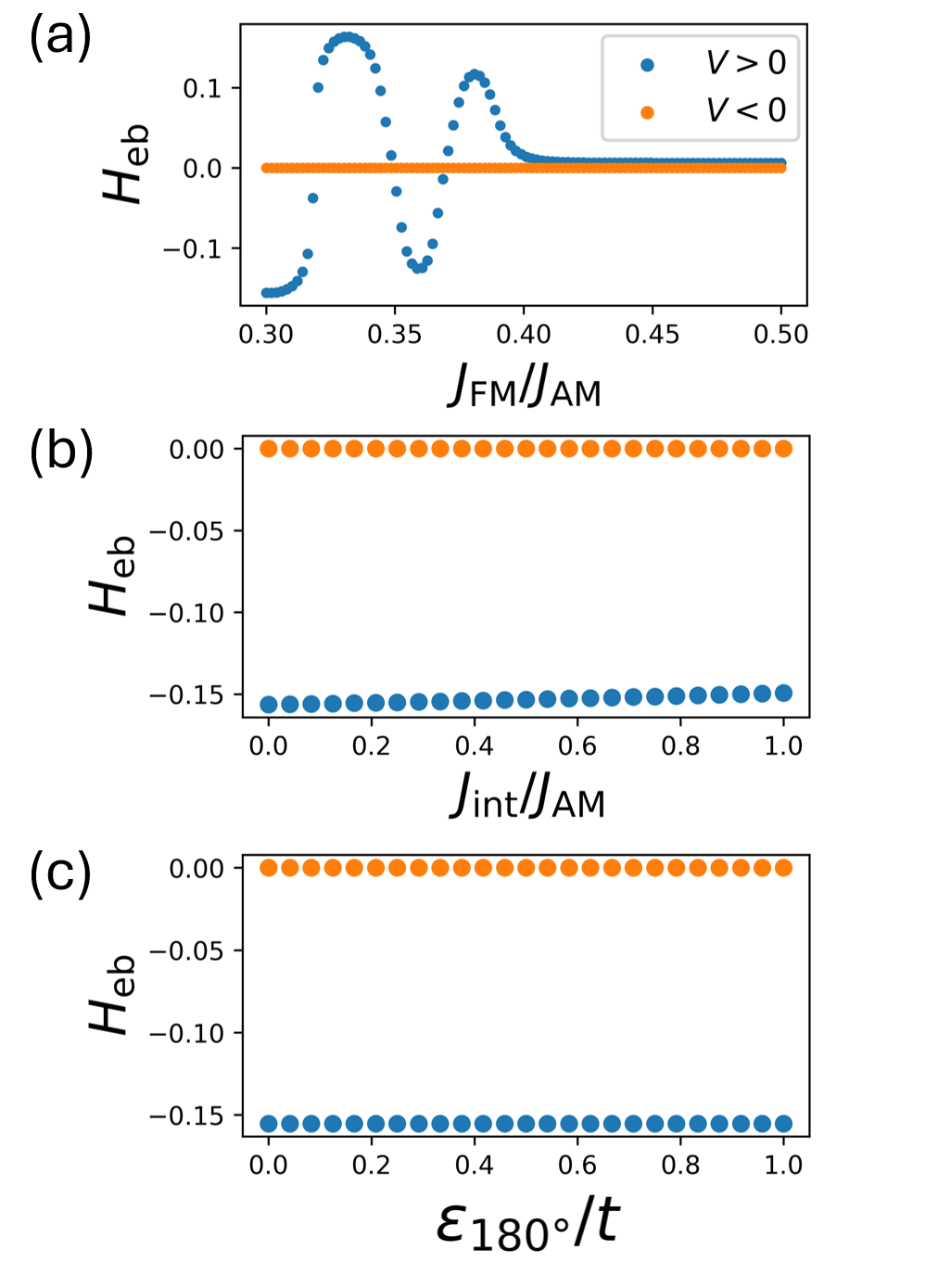}
    \caption{Exchange Bias for a compensated surface, for two cases, the altermagnetic one, that corresponds to a antiferro-orbital-ordering ($V>0$), and the ferro-orbital-ordering ($V<0$) that is more closely related to a normal antiferromagnet. (a) shows the dependence of exchange bias on the stiffness of the ferromagnet. (b) shows the dependence on the interface interaction between the two regions, with $J_{\rm F} /J_{\rm AM} = 0.3$. In (c) we show the dependence on the phenomenological factor of the energy necessary to nucleate the domain wall. }
    \label{fig:compensated}
\end{figure}

\textcolor{black}{
Fig. (\ref{fig:compensated}) shows that, in the altermagnetic case, $H_{\rm eb}$ is nonzero, while in the ferro-orbital-ordering case we have a zero exchange bias. We are not stating that the mechanism of anisotropy generation in all exchange bias systems is due to altermagnetic behaviour, but, in this system, there is such effect.
Additionally, this exchange bias is varely dependent on the exchange intensity of the interface, as seen in Fig. (\ref{fig:compensated}b), thus indicating that the amisotropy is not surface dependent but corresponds to a bulk property. From Fig. (\ref{fig:compensated}a), it should be noted that this phenomenon depends on the stiffness of the FM, so it could pass unnoticed in samples where the magnetic interaction is stronger than the altermagnetic one. The spin split bands are the more likely explanation of the effect in the compensated case. In the direction towards the interface, the different spin species have contrasting Fermi wavelengths that induce an interference pattern of oscillating spin density. This, in turn, contributes to the exchange bias field. By changing $J_{\rm FM}$, the interface conditions of the interference pattern are altered, explaining the oscillation. Finally, Fig. (\ref{fig:compensated}c) shows that the phenomenological parameter $\varepsilon_{180^\circ}$ doesn't have much effect on the simulations, this could well be due to the simplicity of our model or due to using mean field theory methods.
} 

\sectioncustom{Conclusions} Altermagnetic spintronics\cite{DalDin2024, Sun2023, Zhang2024, Guo2023} is in an early stage. Casting new effects and adapting well-known effects to the altermagnetic context might open many scientific and technological possibilities. This work analyzes the behavior of an interface between a ferromagnetic material and an altermagnet. We use a well-established line of arguments and show that emergent phenomena occur because of the altermagnetic materials with spin-split bands. In addition, the model is supplemented by two interaction energies, a Heisenberg-like exchange in spin space $\cH_J$ and an Ising-like exchange in orbital space. We show that the altermagnetic order is impervious to the effects of an external magnetic field for moderated fields.
Furthermore, we anticipate an interfacial exchange interaction induced in the AM/FM interface that manifests itself in the form of a dependence of the energy of the joint system on the angle between the altermagnet \textcolor{black}{orientation} vector and the magnetization. Finally, we report on the effect of exchange bias and discuss its phenomenology, origin \textcolor{black}{and how can it generate anisotropy in fully compensated surfaces}. Exchange bias is usually regarded as a phenomenon observed in magnetic systems composed of a ferromagnetic (FM) layer and an antiferromagnetic (AFM) layer.  We have explored the mechanism behind its altermagnetic version, showing that it exists in principle, and shed some light on the fundamental physics determining its strength. In particular, we have illustrated how electric fields can further manipulate the AM-induced exchange bias effect in a striking difference from the common AFM-based exchange bias bias.
This effect is widely used in various technological applications, particularly in spintronics and magnetic storage. It is fundamental to developing high-performance magnetic devices and ensuring stability, reliability, and efficiency in multiple technologies.



\sectioncustom{Acknowledgements} 
Funding is acknowledged from Fondecyt Regular 1230515.

\bibliography{altermagnets}

\end{document}